\documentclass[aip,cha,author-year,preprint,amsmath,amssymb]{revtex4-1}

\usepackage{graphicx}
\usepackage{dcolumn}
\usepackage{bm}
\usepackage{xspace}
\usepackage{amsmath}
\usepackage{siunitx}
\draft 

\begin{document}


\title{Causal Unit of Rotors in a Cardiac System}
\author{Hiroshi Ashikaga}
\email[]{hashika1@jhmi.edu}
\homepage[]{http://www.hiroshiashikaga.org/}
\affiliation{Cardiac Arrhythmia Service, Johns Hopkins University School of Medicine, 600 N Wolfe Street, Carnegie 568, Baltimore, MD 21287}
\author{Francisco Prieto-Castrillo}
\affiliation{Media Lab, Massachusetts Institute of Technology, Cambridge, Massachusetts, USA}
\author{Mari Kawakatsu}
\affiliation{Program in Applied and Computational Mathematics, Princeton University, Princeton, New Jersey, USA}
\author{Nima Dehghani}
\affiliation{Department of Physics, Massachusetts Institute of Technology, Cambridge, Massachusetts, USA}
\date{\today}

\begin{abstract}
The heart exhibits complex systems behaviors during atrial fibrillation (AF), where the macroscopic collective behavior of the heart causes the microscopic behavior. However, the relationship between the downward causation and scale is nonlinear. We describe rotors in multiple spatiotemporal scales by generating a renormalization group from a numerical model of cardiac excitation, and evaluate the causal architecture of the system by quantifying causal emergence. Causal emergence is an information-theoretic metric that quantifies emergence or reduction between microscopic and macroscopic behaviors of a system by evaluating effective information at each spatiotemporal scale. We find that there is a spatiotemporal scale at which effective information peaks in the cardiac system with rotors. There is a positive correlation between the number of rotors and causal emergence up to the scale of peak causation. In conclusion, one can coarse-grain the cardiac system with rotors to identify a macroscopic scale at which the causal power reaches the maximum. This scale of peak causation should correspond to that of the AF driver, where networks of cardiomyocytes serve as the causal units. Those causal units, if identified, can be reasonable therapeutic targets of clinical intervention to cure AF.

\end{abstract}

\maketitle 

\section{Introduction}

The heart is a multi-scale complex system consisting of five billion autonomous cardiomyocytes, each of which is a nonlinear dynamical system. However, it exhibits relatively simple system behaviors under physiologic conditions. During the regular heart rhythm, the macroscopic behavior of the heart is reducible to microscopic causal behavior and interactions of the cell population in the sinoatrial node (\emph{"supervenience"}). In contrast, complex system behaviors emerge under pathologic conditions where the macroscopic collective behavior of the heart \emph{causes} the microscopic behavior (\emph{"supersedence"}). For example, as soon as the heart undergoes an order-disorder phase transition into fibrillation \citep{ashikaga2017locating}, it controls the behaviors of individual cardiomyocytes to maintain itself. This downward causation is clinically observable in a phenomenon of atrial fibrillation (AF) called \textit{``AF begets AF''}, where a longer duration of pacing-maintained AF results in a longer maintenance of AF after cessation of pacing \citep{wijffels1995atrial}. 

The downward causation from macroscopic to microscopic behaviors of the cardiac system is quantifiable as inter-scale information flow that can be used as a surrogate for the mechanism that maintains AF (\textit{``AF driver''}). In our previous work \citep{ashikaga2017inter}, we demonstate that transfer entropy accurately quantifies the upward and downward information flow between microscopic and macroscopic descriptions of the cardiac system with one of the potential AF drivers, a \textit{rotor}, the rotation center of spiral waves \citep{narayan2012treatment,haissaguerre2014driver,mandapati2000stable}. We have also found that the downward information flow significantly decreases as the description of the system becomes more macroscopic. This subtle but important finding indicates that the relationship between the downward causation and scale is nonlinear. It is possible that, as the system is coarse-grained, it reaches a macroscopic scale at which the causal power peaks. Further coarse-graining removes the fine details of the causal architecture. 

We hypothesize that the cardiac system with rotors has a scale where causal power reaches the maximum. To test the hypothesis, we describe rotors in multiple spatiotemporal scales by generating a renormalization group from a numerical model of cardiac excitation, and evaluate the causal architecture of the system by quantifying \emph{causal emergence ($CE$)}. $CE$ is an information-theoretic metric that quantifies emergence or reduction between microscopic and macroscopic behaviors of a system by evaluating effective information \citep{hoel2013quantifying} at each spatiotemporal scale. Effective information ($EI$) is a quantity that captures causal interactions of a system between its unconstrained repertoire of possible cause and a specific state of possible effect \citep{tononi2003measuring}. 

\section{Methods}

We perform the simulation and the data analysis using Matlab R2016b (Mathworks, Inc.).

\subsection{Model of spiral waves}
We use a monodomain reaction-diffusion model that was originally derived by Fitzhugh~\citep{fitzhugh1961impulses} and Nagumo~\citep{nagumo1962active} as a simplification of the biophysically based Hodgkin-Huxley equations describing current carrying properties of nerve membranes~\citep{hodgkin1952quantitative}, which was later modified by Rogers and McCulloch~\citep{rogers1994collocation} to represent cardiac action potential. This model accurately reproduces several important properties of cardiac systems, including slowed conduction velocity, unidirectional block owing to wavefront curvature, and spiral waves.
\begin{align}
  \frac{\partial v}{\partial t} &= 0.26v(v-0.13)(1-v)-0.1vr+ I_{ex}+\nabla\cdot (D\nabla v)\\
  \frac{\partial r}{\partial t} &= 0.013(v-r)
  \label{eq:FHN02}
\end{align}
Here, $v$ is the transmembrane potential with a finite action potential duration (APD), $r$ is the recovery variable, and $I_{ex}$ is the external current~\citep{pertsov1993spiral}. $D$ is the diffusion tensor, which is a diagonal matrix whose diagonal and off-diagonal elements are 1 mm$^2$/msec and 0 mm$^2$/msec, respectively, to represent a 2-D isotropic system~\citep{rogers1994collocation}. We solved the model equations using a finite difference method for spatial derivatives and explicit Euler integration for time derivatives assuming Neumann boundary conditions. We generate 1,000 sets of a 2-D $120 \times 120$ isotropic lattice of components ($=$ 11.9 cm $\times$ 11.9 cm) by inducing spiral waves with 40 random sequential point stimulations in 40 random components of the lattice (\textit{Supporting Movie 1})\citep{ashikaga2017hidden}. In each component, we computed the time series for 10 seconds excluding the stimulation period with a time step of 0.063 msec, which was subsequently downsampled at a sampling frequency of 400 Hz.

We then defined the instantaneous phase $\phi (t)$ of the time series $s(t)$ in each component via construction of the analytic signal $\xi (t)$, which is a complex function of time.
\begin{equation}
\xi (t)=s(t)+is_H(t)=A(t)e^{i\phi (t)}
\label{eq:analy}
\end{equation}
Here the function $s_H(t)$ is the Hilbert transform of $s(t)$
\begin{equation}
s_H(t)=\frac{1}{\pi}\mathrm{p.v.}\int_{-\infty}^{\infty}\frac{s(\tau)}{t-\tau}d\tau
\label{eq:hilbert}
\end{equation}
where p.v. indicates that the integral is taken in the sense of the Cauchy principal value. We defined the rotor of the spiral wave as a phase singularity \citep{winfree1987time}, where the phase is undefined because all phase values converge. The phase singularity can be localized through calculation of the topological charge $n_t$ \citep{goryachev1996spiral,mermin1979topological}.
\begin{equation}
n_t= \frac{1}{2\pi}\oint_{c} \nabla\phi \cdot d \vec{l}
\label{eq:topch}
\end{equation}
where $\phi (\vec{r})$ is the local phase, and the line integral is taken over the path $\vec{l}$ on a closed curve $c$ surrounding the singularity \citep{bray2002use}. 
\begin{equation}
n_t=\begin{cases}
+1 & \text{ counterclockwise rotor } \\ 
-1 & \text{ clockwise rotor } \\ 
0 & \text{ elsewhere } 
\end{cases}
\label{eq:rotorcharge}
\end{equation}
In this study $|n_t|$ was used to quantify the average number of rotors over the entire time series.

\subsection{Renormalization group}
\begin{figure}[!h]
  \centering
  \includegraphics[width=\linewidth,trim={0cm 3cm 1cm 0cm},clip]{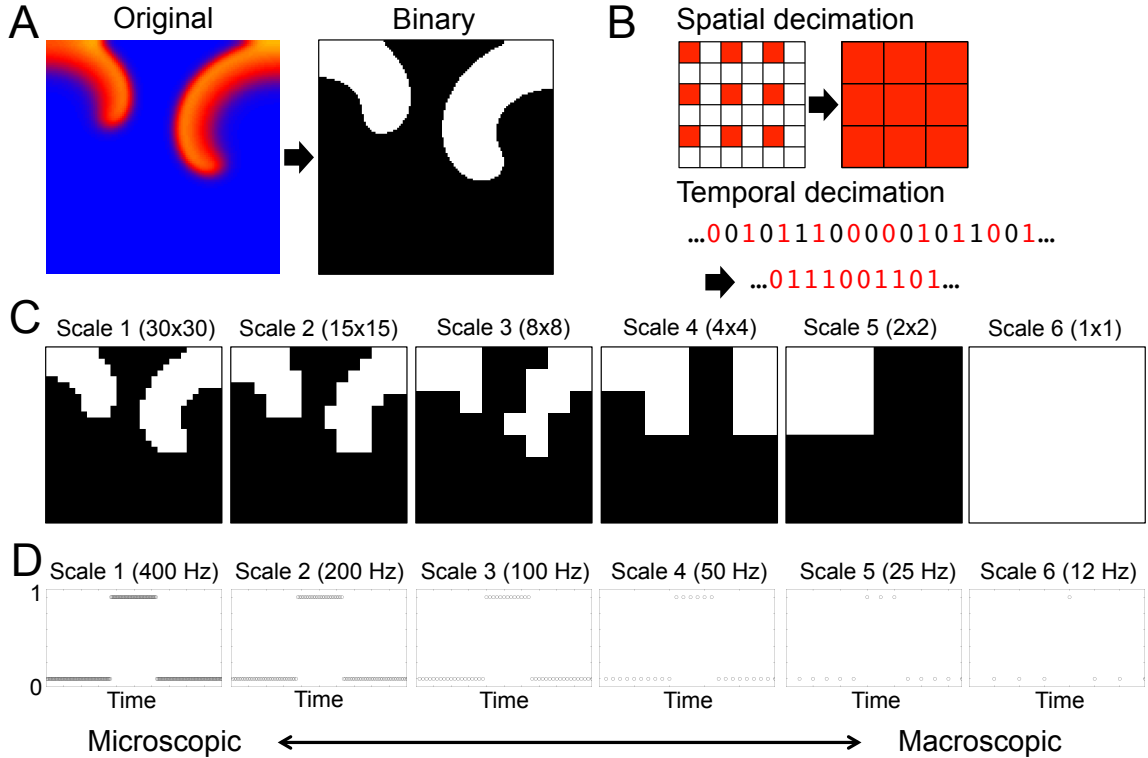}
  \caption{
    \textbf{Renormalization of a cardiac system with spiral waves.} \textit{A. Original description of the system.} For each component, the time series of cardiac excitation is descretized to 1 (black) when excited (during the APD at 90$\%$ repolarization, or APD$_{90}$) or 0 (white) when resting. \textit{B. Spatial and temporal decimation}. Spatial decimation takes the value of cardiac excitation (0 or 1) at each time point in the component at the top left corner of a block of 2 $\times$ 2 immediately adjacent components of the system, and assigns the value to the corresponding site in the system at the next scale. Temporal decimation downsamples the time series of cardiac excitation by a factor of 2. \textit{C. Spatial scales}. Spatial scales include scale 1 (30 $\times$ 30 lattice), scale 2 (15 $\times$ 15 lattice), scale 3 (8 $\times$ 8 lattice), scale 4 (4 $\times$ 4 lattice), scale 5 (2 $\times$ 2 lattice), and scale 6 (1 $\times$ 1 lattice). \textit{D. Temporal scales. } Each circle represents a data sampling point. Temporal scales include scale 1 (400 Hz), scale 2 (200 Hz), scale 3 (100 Hz), scale 4 (50 Hz), scale 5 (25 Hz), and scale 6 (12 Hz).
  }
  \label{fig:renorm}
\end{figure}

We generate a renormalization group of the system by a series of spatial and temporal transformation including coarse-graining and rescaling of the original microscopic description of the system. For each component, the time series of cardiac excitation is descretized to 1 when excited (during the APD at 90$\%$ repolarization, or APD$_{90}$) or 0 when resting (Fig.\ref{fig:renorm}A) \citep{ashikaga2015modelling}. Then we coarse-grain the system spatially and temporally by decimation by a factor of 2 (Figure~\ref{fig:renorm}B). Spatial decimation transforms a $n\times n$ lattice into a $\frac{n}{2} \times \frac{n}{2}$ lattice by extracting the top left component of each 2 $\times$ 2 block (\textit{Supporting Movie 2}). Temporal decimation downsamples the binary time series of each component by a factor of 2. Using a combination of iterative coarse-graining in spatial and temporal axes we create a renormalization group of a total of 36 spatiotemporal scales of the system. The renormalization group includes spatial scales 1 (30$\times$30 lattice), 2 (15$\times$15 lattice), 3 (8$\times$8 lattice), 4 (4$\times$4 lattice), 5 (2$\times$2 lattice), and 6 (1$\times$1 lattice) (Figure~\ref{fig:renorm}C), and temporal scales 1 (400 Hz), 2 (200 Hz), 3 (100 Hz), 4 (50 Hz), 5 (25 Hz) and 6 (12 Hz) (Fig.\ref{fig:renorm}D).

\subsection{Effective information}
We treat each component on the lattice as a time-series process $X$. \textit{Entropy} $H$ of each time-series process $X$ is
\begin{equation}
H(X)=-\sum_{x}p(x)\log_{2}p(x)
\label{eq:entropy}
\end{equation}
where $p(x)$ denotes the probability density function of the time series generated by $X$.
\textit{Effective information} quantifies the information generated when the system enters a specific state of possible effect $Y$ out of its unconstrained probability distribution of possible cause $X$ \citep{tononi2003measuring}.
\begin{eqnarray}
EI(X\rightarrow Y)&=&I(X;Y)\\
&=&H(X)+H(Y)-H(X,Y)\\
&=&\sum_{x,y}p(x,y)\log_{2}\frac{p(x,y)}{p(x)p(y)}
\label{eq:mi}
\end{eqnarray}
where $X$ has a uniform probability distribution so that it provides the maximum entropy $H(X)_{max}$. $I(X;Y)$ is mutual information, $p(x,y)$ and $H(X,Y)$ denote the joint probability density function and the joint entropy of $X$ and $Y$, respectively. Mutual information is originally a measure of statistical dependence to quantify how much information is shared between a source and a destination \citep{shannon1948mathematical}. In this context, however, mutual information is applied between two time series of a system that is first perturbed into all possible states with equal probability and then observed as a sepcific state. Because of the system perturbations, mutual information here is a causal measure, and thus effective information of the system is a state-independent informational measure of a system’s causal architecture \citep{hoel2013quantifying}.

\begin{figure}[!h]
  \centering
  \includegraphics[width=\linewidth,trim={0cm 7cm 1cm 0cm},clip]{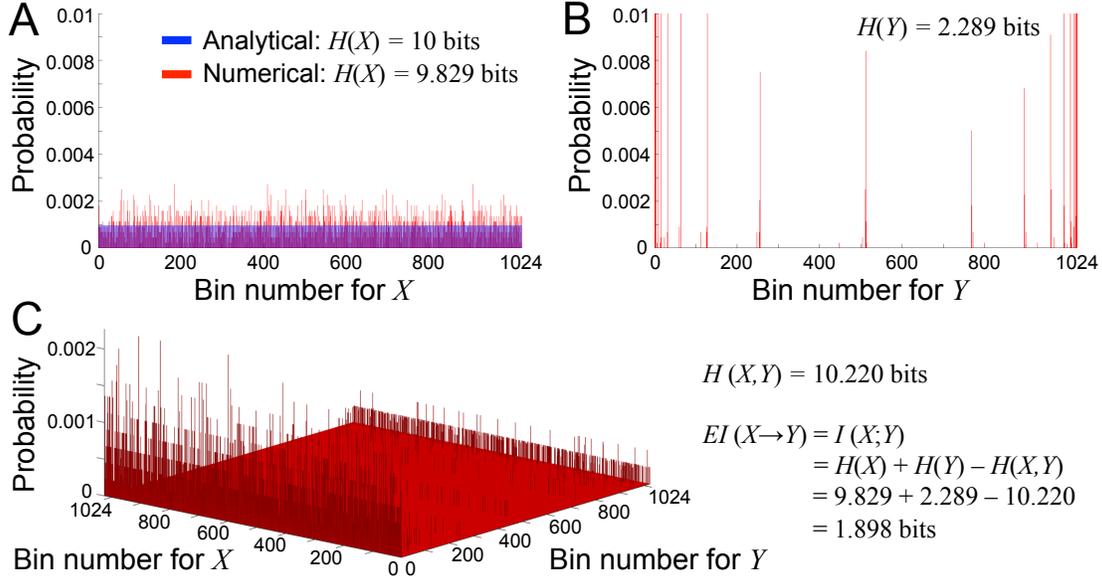}
  \caption{
    \textbf{Probability distribution of cause $X$ and effect $Y$.} We define the bin number $b=2^{10}$ in this study. \textit{A. Unconstrained probability distribution of possible cause $X$.} Analytically, the probability of all bins is uniformly $1/b$ (shown in blue), and thus entropy is equal to the maximum entropy at $\log_{2}b$=10 bits. In contrast, numerically, the probability is non-uniform due to the discretization effect (shown in red). Entropy is 9.829 bits, which close to but not identical to the maximum entropy. \textit{B. Probability distribution of a specific state of possible effect $Y$}. The probability is non-uniform. Entropy is 2.289 bits in this case. \textit{C. Bivariate probability distribution of cause $X$ and effect $Y$}. Joint entropy is 10.220 bits in this case. Effective information from case $X$ to effect $Y$ is equal to mutual information between $X$ and $Y$, thus is calculated as 1.898 bits.
  }
  \label{fig:prob}
\end{figure}

One can describe a $n \times n$ lattice at time $t$ as a binary string of length $n \times n$. Therefore, the unconstrained repertoire of all possible causes $X$ at time $t_0$ consists of $2^{n^2}$ possible states with equal probability ${1/{2^{n^2}}}$ at each time point. We define the bin number $b$ ($b<2^{n^2}$) to calculate the probability distribution of $X$ and $Y$, and we use $b=2^{10}=1,024$ in this study. Analytically, because $X$ has a uniform probability distribution, the probability that $X$ falls in one of the $b$ bins at each time point is $1/b$. Therefore, entropy of $X$ is equal to the maximum entropy (Figure~\ref{fig:prob}A).
\begin{eqnarray}
H(X)&=&-\sum_{x}p(x)\log_{2}p(x)\\
&=&b\times (-\frac{1}{b}\log_{2}\frac{1}{b})\\
&=&\log_{2}b
\label{eq:entropy}
\end{eqnarray}
Numerically, $X$ can be defined as a vector of uniformly distributed random numbers between 1 and $2^{n^2}$-1 for a time series of finite duration. Due to the discretization effect, the probability is non-uniform. Entropy is close to but not identical to the maximum entropy (Fig.\ref{fig:prob}A). Similarly, $Y$ can be defined as a vector of decimal numbers between 1 and $2^{n^2}$-1, each of which represents a specific state of the system with rotors (Fig.\ref{fig:prob}B). 

\textit{Causal emergence} is a difference in effective information between scales.
\begin{equation}
CE=EI(X_{m}\rightarrow Y_{m})-EI(X_{n}\rightarrow Y_{n})
\label{eq:ei01}
\end{equation}
where $m$ and $n$ are different scales of the system description from the renormalization group. When scale $m$ is more macroscopic than scale $n (m>n)$, a positive $CE$ indicates that the macroscopic behavior is emergence (supersedence), whereas a negative $CE$ indicates that the macroscopic behavior is reduction (supervenience) \citep{hoel2013quantifying}. In this study we quantify causal emergence with respect to the most microscopic system description with spatial scale = temporal scale = 1.

\section{Results}
\subsection{Evaluation of variance of effective information to quantify rotor dynamics}
\begin{figure}[!h]
  \centering
  \includegraphics[width=0.8\linewidth,trim={0cm 10cm 0cm 0cm},clip]{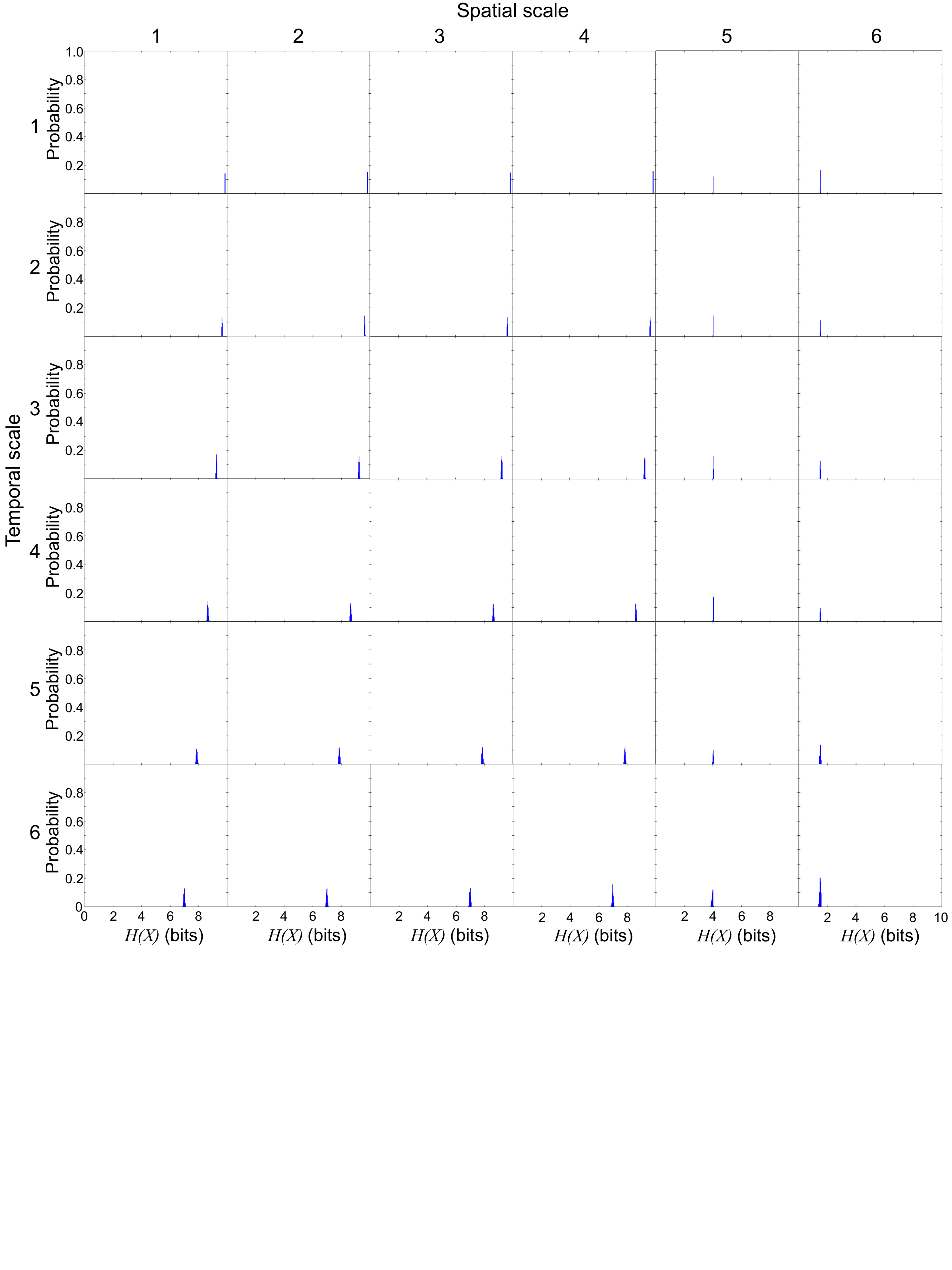}
  \caption{
    \textbf{Entropy of unconstrained probability distribution of possible cause $X$ in a representative spiral wave data set.} $H(X)$ is not uniquely determined due to the discretization effect, but the variance is small. Each subplot represents he probability distribution of $H(X)$. The columns represent the spatial scales (1 through 6) and the rows represent the temporal scales (1 through 6). 
  }
  \label{fig:huc0}
\end{figure}

\begin{figure}[!h]
  \centering
  \includegraphics[width=0.8\linewidth,trim={0cm 10cm 0cm 0cm},clip]{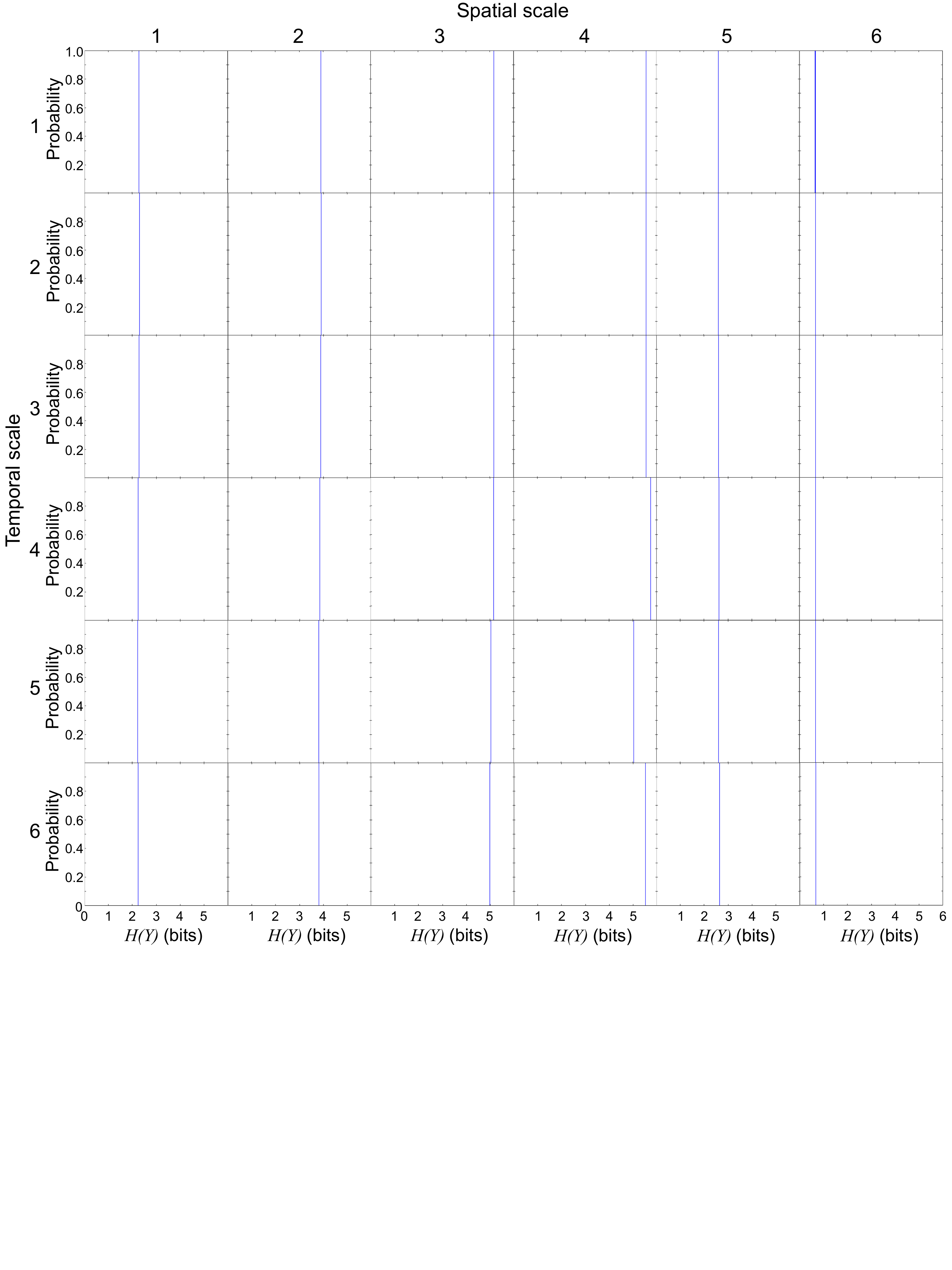}
  \caption{
    \textbf{Entropy of specific state of possible effect $Y$ in a representative spiral wave data set.} $H(Y)$ is uniquely determined because it represents a specific state of the system regardless of the spatiotemporal scale. Each subplot represents the probability distribution of $H(Y)$. The columns represent the spatial scales (1 through 6) and the rows represent the temporal scales (1 through 6). 
  }
  \label{fig:hue0}
\end{figure}

\begin{figure}[!h]
  \centering
  \includegraphics[width=0.8\linewidth,trim={0cm 10cm 0cm 0cm},clip]{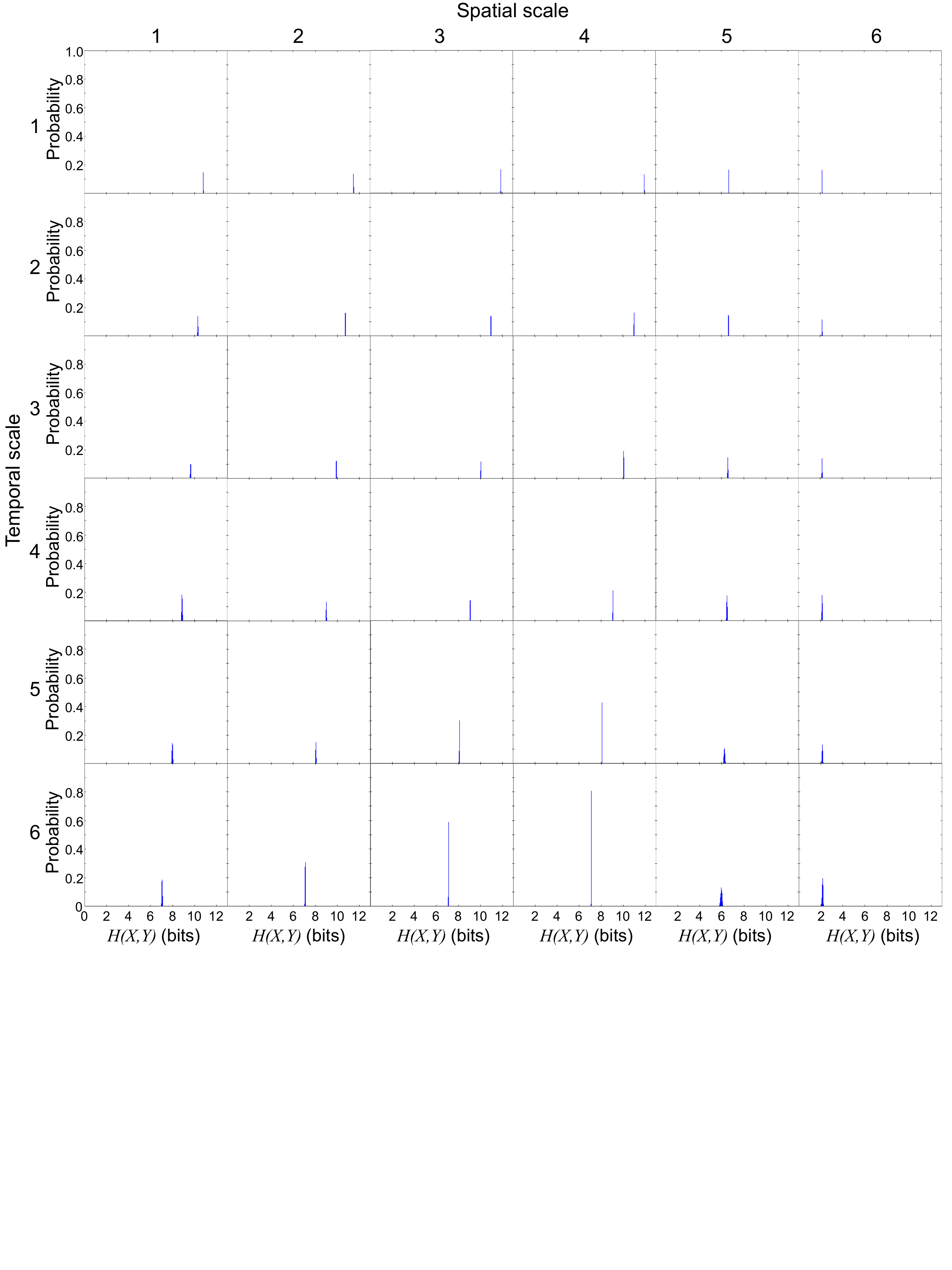}
  \caption{
    \textbf{Joint entropy of cause $X$ and effect $Y$ in a representative spiral wave data set.} $H(X,Y)$ is not uniquely determined due to the discretization effect, but the variance is small. Each subplot represents the probability distribution of $H(X,Y)$. The columns represent the spatial scales (1 through 6) and the rows represent the temporal scales (1 through 6). 
  }
  \label{fig:huce0}
\end{figure}

\begin{figure}[!h]
  \centering
  \includegraphics[width=0.8\linewidth,trim={0cm 10cm 0cm 0cm},clip]{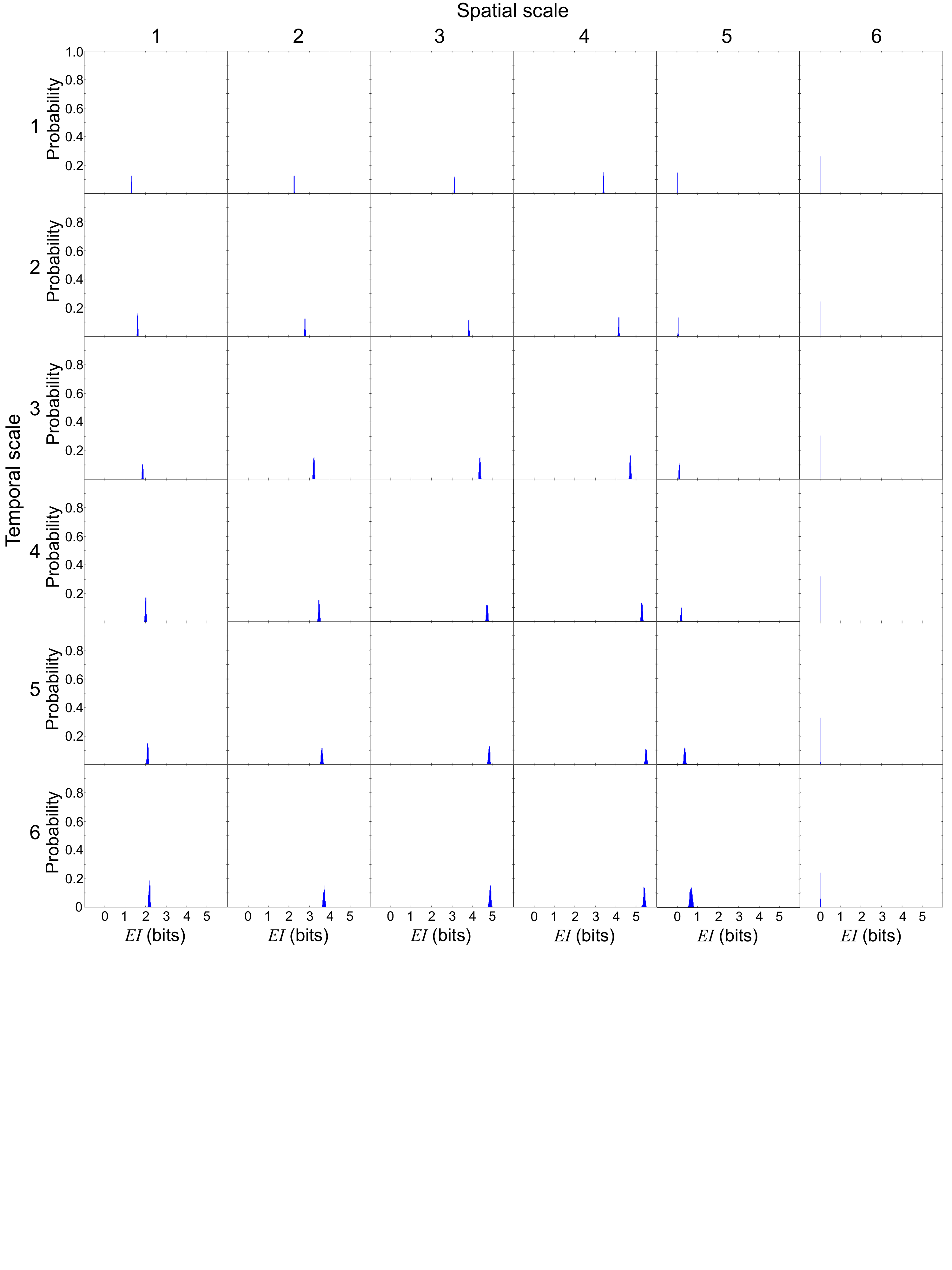}
  \caption{
    \textbf{Effective information from cause $X$ to effect $Y$ in a representative spiral wave data set.} $EI(X\rightarrow Y)[=I(X;Y)=H(X)+H(Y)-H(X,Y)]$ is not uniquely determined due to the discretization effect, but the variance is small. Each subplot represents the probability distribution of $EI(X\rightarrow Y)$. The columns represent the spatial scales (1 through 6) and the rows represent the temporal scales (1 through 6). 
  }
  \label{fig:ei0}
\end{figure}

First we evaluate the variance of effective information in rotor dynamics. We repeat 1,000 numerical computations of $X$ and $Y$ in a representative spiral wave data set to calculate entropy $H(X)$, $H(Y)$, $H(X,Y)$, then calculate $EI(X\rightarrow Y)$. Numerically, $H(X)$ is not uniquely determined due to the discretization effect, but the variance is small (Fig.\ref{fig:huc0}). spatial coarse-graining has minimal impact on the probability ditribution of $H(X)$ from scales 1 through 4, but $H(X)$ steeply falls in scales 5 and 6. In contrast, temporal coarse-graining gradually shifts the distribution of $H(X)$ to the left. $H(Y)$ is uniquely determined because it represents a specific state of the system rgardless of the spatiotemporal scale (Fig.\ref{fig:hue0}). In this case, spatial coarse-graining clearly increases the distribution of $H(Y)$ to the right, which peaks at scale 4 and decreases at scales 5 and 6. Similarly, temporal coarse-graining increases the distribution of $H(Y)$ to the right, which peaks at scale 4 and decreases at scales 5 and 6. The relationship between the spatiotemporal coarse-graining and the probability distribution of joint entropy $H(X,Y)$ is similar to that of $H(X)$ (Fig.\ref{fig:huce0}), and the variance remains small. Effective infromation $EI(X\rightarrow Y)$ peaks at spatial scale of 4 and temporal scale 5, and the variance of $EI(X\rightarrow Y)$ remains small (Fig.\ref{fig:ei0}). This result indicates that, despite the discretization effect, numerical computation of $EI(X\rightarrow Y)$ is robust with a high reproducibility, and thus $EI(X\rightarrow Y)$ can be used to quantify the information of rotor dynamics at each spatiotemporal scale. 

\subsection{Evaluation of effective information in aggregate data sets}
\begin{figure}[!h]
  \centering
  \includegraphics[width=0.6\linewidth,trim={0cm 1cm 13cm 0cm},clip]{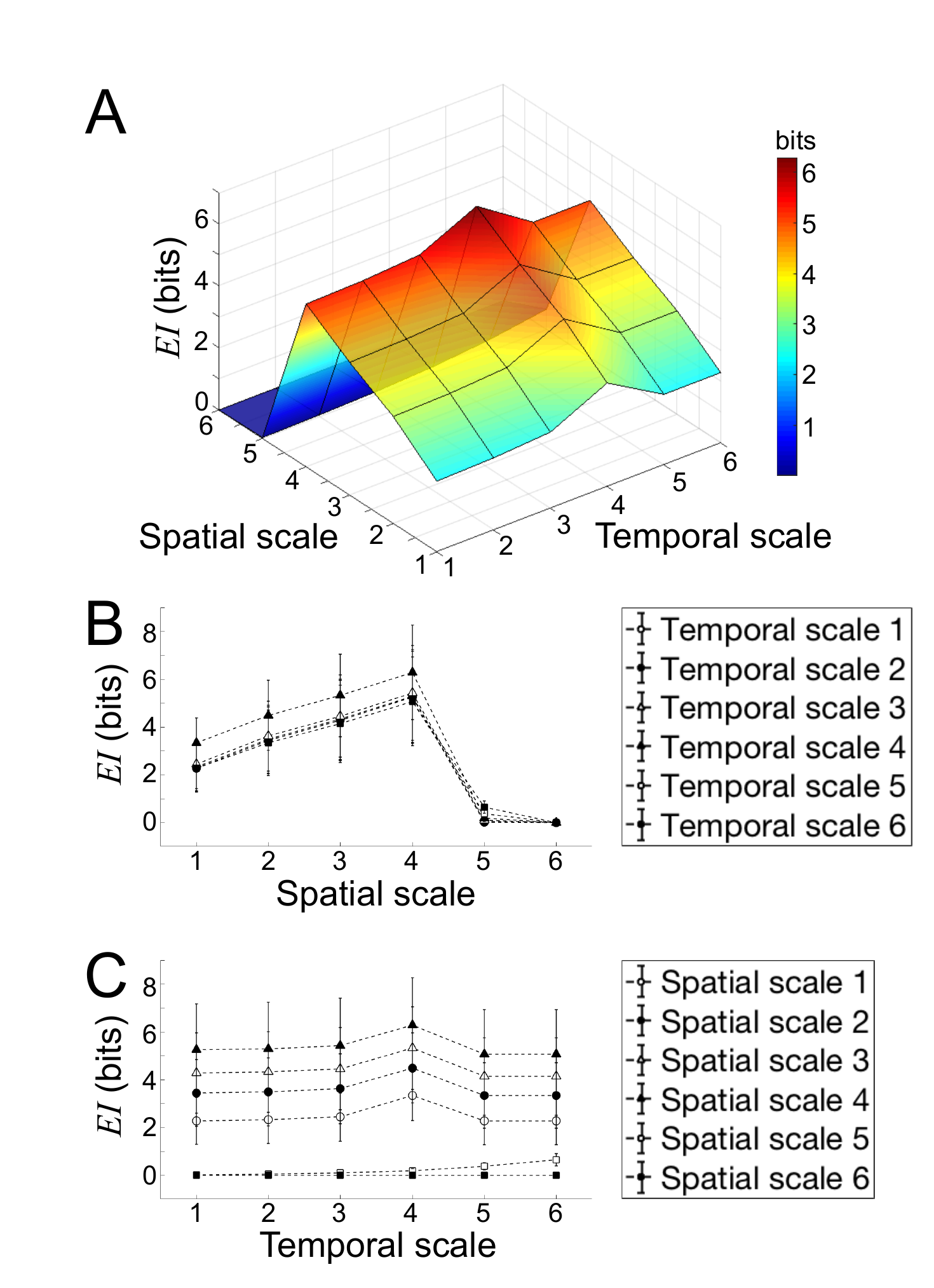}
  \caption{
    \textbf{Effective information of the system in aggregate data sets.} \textit{A. Overview.} Each point indicates the mean effective information $EI(X\rightarrow Y)$ of 1,000 data sets at each spatiotemporal scale. $EI(X\rightarrow Y)$ reaches the global maximum at spatial scale = temporal scale = 4. \textit{B. Effective information \textit{vs.} spatial scale}. \textit{C. Effective information \textit{vs.} temporal}. Each point indicates the mean of $EI$ of 1,000 data sets at each spatiotemporal scale).
  }
  \label{fig:ei}
\end{figure}

Next, we quantify effective information of the renormalization group of a total of 36 spatiotemporal scales of the system in aggregate data of 1,000 sets (Fig.\ref{fig:ei}). Overall, effective information increases as the scale increases from microscopic to macroscopic descriptions of the system. However, it reaches the global maximum at spatial scale = temporal scale = 4, beyond which effective information decreases (Fig.\ref{fig:ei}). The difference in effective information between scales is larger in spatial coarse-graining (Fig.\ref{fig:ei}B) than that of temporal coarse-graining (Fig.\ref{fig:ei}C), indicating that the impact of spatial coarse-graining on effective information is higher than that of temporal coarse-graining. 
 
\subsection{Relationship between the number of rotors and causal emergence}
\begin{figure}[!h]
  \centering
  \includegraphics[width=0.4\linewidth,trim={1cm 13cm 19cm 0cm},clip]{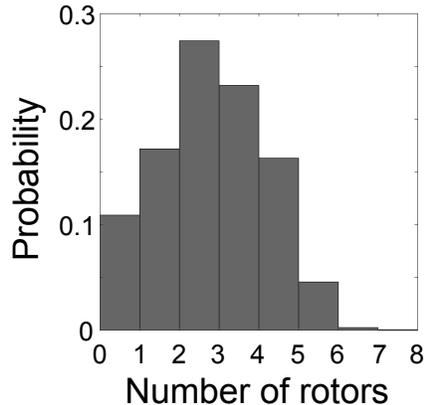}
  \caption{
    \textbf{Probability distribution of the number of rotors.} The number of rotors ranges from 0 to 7 in 1,000 data sets. 
  }
  \label{fig:rotors}
\end{figure}

\begin{figure}[!h]
  \centering
  \includegraphics[width=0.8\linewidth,trim={0cm 10cm 0cm 0cm},clip]{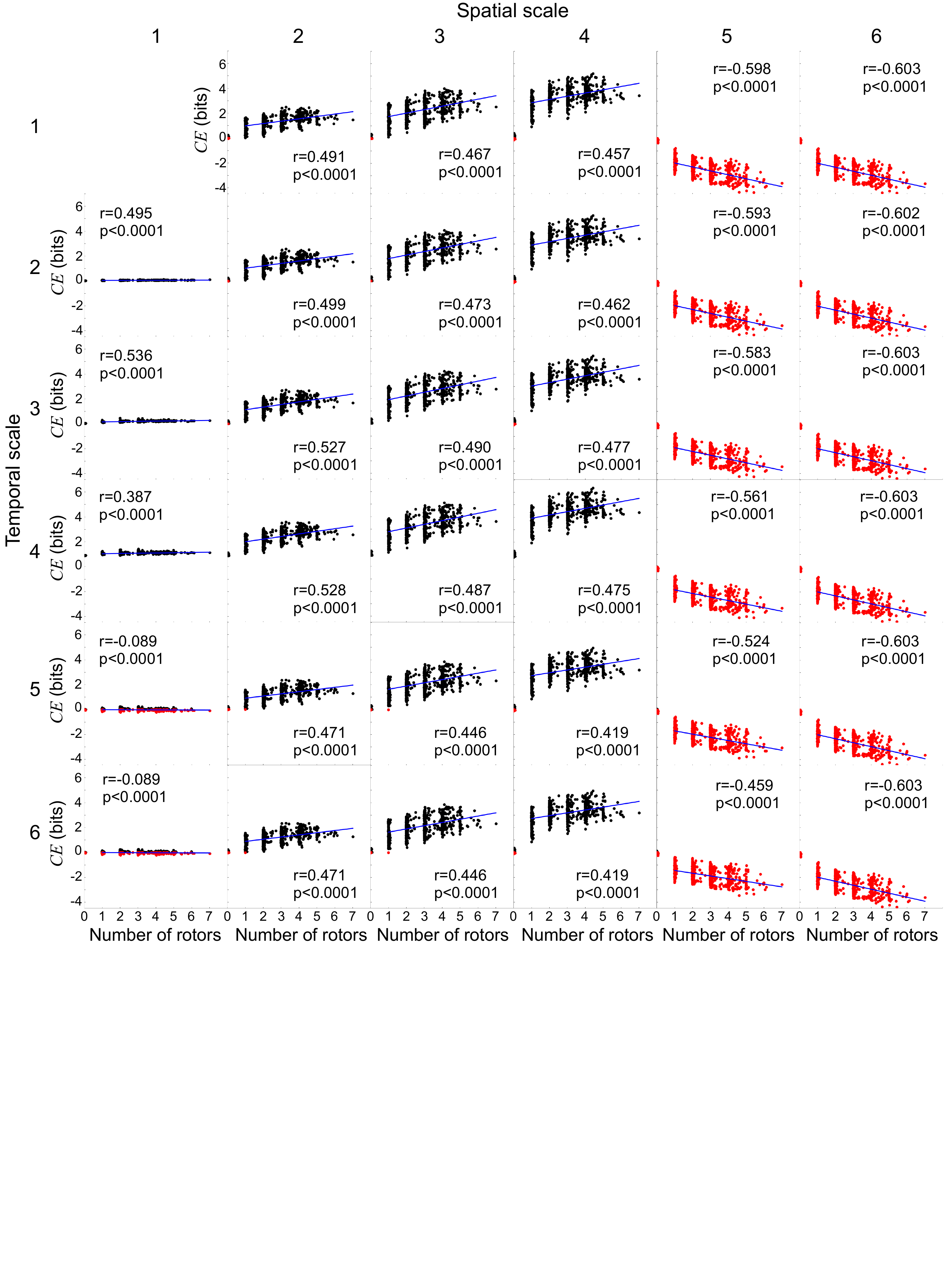}
  \caption{
    \textbf{Number of rotors and causal emergence.} We quantify causal emergence ($CE$) with respect to the most microscopic system description at spatial scale = temporal scale = 1. Each subplot represents an association between the number of rotors and $CE$ at each spatiotemporal scale. Black dots indicate $CE>$0 (emergence), whereas red dots indicate $CE<$0 (reduction). Blue lines indicate linear fit for the number of rotors $\geq$1. The columns represent the spatial scales (1 through 6) and the rows represent the temporal scales (1 through 6). 
  }
  \label{fig:ce}
\end{figure}

Lastly, we evaluate the relationship between the number of rotors and causal emergence in individual data sets. The number of rotors ranges from 0 to 7, with a median of 3 (Fig.\ref{fig:rotors}). For system descriptions at spatial scale$\leq$4 and temporal scale$\leq$4, causal emergence is positive for all the data sets except a few where a rotor prematurely disappears on its own (number of rotors$\leq$1, red dots in Fig.\ref{fig:ce}). There is a significant positive correlation between the number of rotors and causal emergence, indicating that causal emergence consistently increases as the number of rotors increases. For system descriptions at spatial scale$\geq$5, causal emergence is negative for all the data sets, and there is a significant negative correlation between the number of rotors and causal emergence. This indicates that the macroscopic behavior at those scales are reducible to the microscopic behavior. For system descriptions at spatial scale=1 and temporal scale$\geq$5, causal emergence scatters in positive and negative values. This indicates that the causal relationship at those scales is inconsistent. There is still a significant negative correlation between the number of rotors and causal emergence, but the correlation coefficients are small. For system descriptions at spatial scale=2,3,4 and temporal scale$\geq$5, causal emergence is almost always positive and there is a significant positive correlation between the number of rotors and causal emergence. This result indicates that temporal coarse-graining has a smaller impact than spatial coarse-graining on the causal architecture.

\section{Discussion}

\subsection{Main findings}
First, we find that numerical computation of effective information in the cardiac system with rotors is robust with high reproducibility despite the discretization effect. Therefore effective information can be used to quantify the information of rotor dynamics at each spatiotemporal scale. 

Next, we find that there is a spatiotemporal scale at which effective information peaks in the cardiac system with rotors. This suggests the presence of causal units at this scale consisting of networks of components that serve as the AF driver.

Lastly, we find that there is a positive correlation between the number of rotors and causal emergence up to the scale of peak causation. This indicates that the causal relationship between the macroscopic and the microscopic behaviors reverses beyond that scale.

\subsection{Causal architecture of rotors}
AF currently impacts the lives of 33 million patients worldwide\citep{chugh2013worldwide,rahman2014global}. Importantly, AF is associated with a five-fold increased risk of thromboembolism such as stroke \citep{wolf1991atrial}, and accounts for 15$\%$ of strokes overall \citep{atrial1994risk}. In addition, AF increases the risk of cognitive impairment \citep{kalantarian2013cognitive,thacker2013atrial} independent of clinical stroke. AF is also associated with a 2-fold increased risk of dementia \citep{ott1997atrial}, and more than 10$\%$ of AF patients develop dementia over 5 years \citep{miyasaka2007risk}. Furthermore, AF is a powerful risk factor of myocardial infarction \citep{soliman2014atrial} and death \citep{benjamin1998impact}. Although the mechanism that initiates AF is ascribed to focal triggers primarily from the pulmonary veins \citep{haissaguerre1998spontaneous}, the AF driver remains unknown. 

In this study we describe rotors in multiple spatiotemporal scales by generating a renormalization group of the cardiac system and evaluate the causal architecture of the system by quantifying causal emergence. Causal emergence was originally developed in neuroscience but is applicable to any multi-scale systems \citep{hoel2017map}. Our analysis using causal emergence confirms that rotors are emergent behaviors of the heart, that is, macroscopic collective behaviors that \textit{cause} microscopic behaviors. This indicates that a multi-scale, complex systems approach is an appropriate direction of investigation to understand the AF driver, rather than the reductionistic approach to understanding the AF driver by describing microscopic behaviors of the system with near-infinite degrees of freedom.

In this particular cardiac system, we find that effective information peaks at spatial scale = temporal scale = 4. The spatial scale of 4 divides the original 11.9cm $\times$ 11.9cm lattice of cardiac system into 4$\times$4 units, each of which occupies 3cm$\times$3cm in size. This 2-D unit contains 5-7.5$\times 10^5$ cells with dimensions of human atrial cardiomyocytes (\textit{e.g.}, \SI{120}{\micro\metre} length and 10-15\SI{}{\micro\metre} diameter \citep{nygren1998mathematical}). It is likely that different cardiac systems and patients have different sizes of causal unit of rotors just as the ``critical mass'' needed to sustain fibrillation  \citep{mcwilliam1887fibrillar,garrey1914nature} is different for different patients (``effective size'' \citep{panfilov2006heart}). However, the important point is that one can coarse-grain the cardiac system with rotors to identify a macroscopic scale at which the causal power reaches the maximum. This scale of peak causation should correspond to that of the AF driver, where networks of cardiomyocytes serve as the causal units.

\subsection{Limitations}
We used a modified Fitzhugh-Nagumo model, which is a relatively simple model of excitable media, with a homogeneous and isotropic lattice. It is possible that our findings may not directly be extrapolated to a more realistic cardiac system with tissue heterogeneity and anisotropy. However, the information-theoretic metrics in this study are independent of any specific trajectory of each rotor. Therefore, our approach is applicable to any other cardiac system.

\subsection{Conclusions}
One can coarse-grain the cardiac system with rotors to identify a macroscopic scale at which the causal power reaches the maximum. This scale of peak causation should correspond to that of the AF driver, where networks of cardiomyocytes serve as the causal units. Those causal units, if identified, can be reasonable therapeutic targets of clinical intervention to cure AF.

\section*{Funding}
This work was supported by the Fondation Leducq Transatlantic Network of Excellence.

\section*{Supplementary material}

\paragraph*{Supporting Movie 1.}
\label{sample_stim_movie}
\textit{Random sequential point stimulations.} We induce spiral waves by introducing 40 random sequential point stimulations in 40 random components of the lattice. In this example, random sequential point stimulations induce five spiral waves.

\paragraph*{Supporting Movie 2.}
\label{renorm}
\textit{Renormalzation group.} The movie shows a renormalization group of the cardiac system with two spiral waves by a series of transformation including coarse-graining and length rescaling (scale 1 through 6). For each component, the time series of cardiac excitation is descretized to 1 (black) when excited (during the APD at 90$\%$ repolarization, or APD$_{90}$) or 0 (white) when resting.

\bibliography{ce}

\end{document}